\documentstyle[12pt]{article}

\begin{document}
\begin{center}
{\Large \bf Entanglement of neutrino states}

\bigskip

{\large D.L.~Khokhlov}
\smallskip

{\it Sumy State University, Ukraine\\
E-mail: dlkhokhl@rambler.ru}

\end{center}

\begin{abstract}
Muon and muon antineutrino born in the decay of charged pion form
the entangled spin state. The decay of muon with the left helicity
triggers the left helicity for muon antineutrino to preserve the
null total angular momentum of muon and muon antineutrino. This is
forbidden for antineutrino hence one cannot detect the muon
antineutrino after the decay of muon. This effect may explain the
deficit of muon neutrino flux in the Super-Kamiokande, K2K, MINOS
experiments.

\end{abstract}

\noindent
Key words: entanglement; decay of charged pion; atmospheric and
accelerator neutrinos

\bigskip

As known quantum mechanics admits the existence of the global
state for a number of particles, so called entangled state, which
cannot be written as a product of the individual states of the
particles. The phenomenon of entanglement was introduced by
Einstein, Podolsky, and Rosen and then was confirmed in Bell
experiments through the quantum correlations between entangled
states violating Bell inequalities, e.g.~\cite{Shi} and references
therein. Entangled state may arise for the spin states of
the particles outgoing some reaction. Then, the total angular
momentum of the particles outgoing the reaction maintains
preserved. Detection of one of the particles in a certain spin
state triggers the spin states of the other
particles to preserve the total angular momentum.
We shall study entanglement in the decay of charged pion.

Consider decay of charged pion~\cite{Co}
\begin{equation}
\pi^{-}\rightarrow\mu^{-}+\bar{\nu}_{\mu}.
\label{eq:pi}
\end{equation}
Muon in turn decays as
\begin{equation}
\mu^{-}\rightarrow{e}^{-}+\bar{\nu}_{e}\nu_{\mu}.
\label{eq:mu}
\end{equation}
The weak interaction prefers left-hand particles and right-hand
antiparticles. By virtue of Lorentz invariance massive particle in
the rest frame may have both the left and right helicity with
equal probability. For the moving particle the probability of
having left helicity predominates and for the moving antiparticle
the probability of having right helicity predominates.
While the velocity of the particle reaches the velocity of light,
$v\rightarrow c$, the probability of having left
helicity for particle and right helicity for antiparticle tends to
unity. Neutrino propagating with the velocity of light has the
left helicity while antineutrino has the right helicity.
We suppose that neutrinos have zero mass.

The end products of the decay of pion, $\mu^-$ and
$\bar{\nu}_{\mu}$, form the entangled spin state
\begin{equation}
|\psi\rangle=\frac{1}{\sqrt{2}}
(|\bar{\nu}\rangle_1|\mu^-\rangle_2-
|\bar{\nu}\rangle_2|\mu^-\rangle_1)
\label{eq:psi}
\end{equation}
where indices $1$ and $2$ label spin states which are identical.
The total angular momentum of muon and muon antineutrino is equal to
the angular momentum of pion, which is zero.
In the system of pion, muon antineutrino is born in the state
$|\bar\nu_\mu\rangle=|-k,+1/2\rangle$,
and muon in the state
$|\mu^-\rangle=|k,+1/2\rangle$
where $k$ stands for the momentum in the system of pion,
$1/2$ stands for the helicity.
Then, the total spin of the system of muon antineutrino
and muon defined onto a certain momentum is equal to zero.

Consider decay of muon eq.~(\ref{eq:mu}).
Since electron is preferably born as a left-hand particle, muon
decays as a left-hand particle, accounting for the total spin of
electron antineutrino and muon neutrino is equal to zero.
Hence, muon emerges as a right-hand particle
and decays as a left-hand particle.
By virtue of Lorentz invariance electron born in the decay of muon
has a small probability to have right helicity. The fraction
of right-hand electrons grows with the decrease of energy. We
shall omit this effect as small to change the result
significantly.
The decay of muon leads to the collapse of the global wave
function of $\mu^-$ and $\bar{\nu}_{\mu}$ eq.~(\ref{eq:psi}).
Muon decays into the state $|\mu^-\rangle=|k,-1/2\rangle$ that
must trigger the state of muon antineutrino
$|\bar\nu_\mu\rangle=|-k,-1/2\rangle$ to preserve the null total
angular moment.
This is forbidden because antineutrino has the right helicity.
Within the lifetime of muon one can detect the muon antineutrino.
After the decay of muon one cannot detect the muon antineutrino.

The Super-Kamiokande experiment~\cite{SK}
detected atmospheric electron and
muon neutrinos and their antineutrinos which are produced in the
hadronic showers induced by primary cosmic rays in the earth's
atmosphere.
The production of atmospheric neutrinos occurs in the decays
of charged pions, $\pi^\mp$, (K-mesons, $K^\mp$).
The ratio for the atmospheric muon and electron neutrino fluxes is
given by
$R\equiv[(\nu_\mu+\bar{\nu}_\mu)/(\nu_e+\bar{\nu}_e]$.
The calculated ratio is $R(calc)\sim 2$.
The data of the Super-Kamiokande experiment~\cite{SK} yields the
double ratio depending on the energies of neutrinos
\begin{eqnarray}
R(data/calc)=0.63\pm 0.03\,(stat.)\pm 0.05\,(syst.)
\quad E<1.33\ {\mathrm GeV}\\
\nonumber
R(data/calc)=0.65\pm 0.05\,(stat.)\pm 0.08\,(syst.)
\quad E>1.33\ {\mathrm GeV}
\label{eq:RSK}
\end{eqnarray}

The distances $L$ traveled by neutrinos before they reach the
detector vary in a wide range: for vertically downward going
neutrinos (neutrino zenith angle $\Theta_\nu=0$) $L\sim 15$ km,
for horizontal neutrino trajectories ($\Theta_\nu=90^\circ$)
$L\sim 500$ km, for the vertically up-going neutrinos
($\Theta_\nu=180^\circ$) $L\sim 13\,000$ km.
The $\mu$-like data exhibit a strong asymmetry in zenith angle
while e-like data exhibit no significant asymmetry. Asymmetry is
defined as $A\equiv (U-D)/(U+D)$ where $U$ is the number of
upward going events with $-1<\cos\Theta <-0.2$ and $D$ is the
number of downward going events with $0.2<\cos\Theta <1$.
The flux of atmospheric neutrinos is
expected to be nearly up-down symmetric for neutrino energies
$E>1.33$ GeV above which effects due to the Earth's magnetic
field on cosmic rays are small.
The data exhibit the asymmetry $A\simeq 0$ for e-like events for
all momenta. For $\mu$-like events the asymmetry is close
to zero at low momenta and decreases with momentum.
For multi-GeV $\mu$-like events ($E> 1.33$ GeV) the measured
asymmetry is $A=-0.296\pm 0.048\pm 0.01$~\cite{SK}.

The lifetime of muon is $\tau_\mu=2.2\times 10^{-6}$ s~\cite{Gr}.
The lifetime of muon in the laboratory frame is multiplied by the
Lorentz factor. Take muon neutrinos and antineutrinos born in the
decay of pion of the energy $E=1.33$ GeV and correspondingly
muons born in the decay of pion of the energy $E=1.33$ GeV.
This yields the Lorentz factor of muon
$\gamma=12.6$. Then the travel path of muon
in the laboratory frame is $L_0=c\gamma\tau_\mu=830$ km.
This is more than the distances traveled by vertically downward
and horizontal going neutrinos but less than the distance traveled
by vertically upward going neutrinos.
Muons of the energy $E>1.33$ GeV decay before detecting upward
but after downward going muon neutrinos and antineutrinos of the
energy $E>1.33$ GeV born in the decay of pion.
Due to the entanglement of muon and neutrino states
one can detect downward but cannot upward going muon neutrinos and
antineutrinos born in the decay of pion.
Then, due to the entanglement of muon and neutrino states
the asymmetry for $\mu$-like events is expected to be
$A\sim -1/3$ that is in agreement with the measured value.
Thus, the entanglement of muon and neutrino states may explain
the deficit of upward going muon neutrinos of the energy $E>1.33$
GeV in the Super-Kamiokande experiment.
The lifetime of muon in the laboratory frame decreases with the
decrease of the energy of muon. This may explain reduction of the
asymmetry for $\mu$-like events $E<1.33$.

The K2K experiment~\cite{K2K} detected muon neutrinos from the
KEK-PS beam in two detectors located along the beam axis at
distances of 300 m (Near Detector) and 250 km
(Far Detector). It is measured the rate of $\nu_\mu$
born in the decay of $\pi^+$.
The neutrino energy distribution is centered on $E\sim 1.3$ GeV.
112 events are observed at the Far Detector compared to an
expectation of $158.1+9.2-8.6$ events.

The deficit of muon neutrino flux in the K2K experiment may be
accounted for by the entanglement of muon and neutrino
states. After the decay of muon one cannot detect muon neutrino.
We shall estimate the number of undecayed muons for the time
needed for muon neutrino to travel from the accelerator to the Far
Detector.
Estimate the effect for the peak neutrino energy $E\sim 1.3$ GeV
hence for the muon energy $E=1.3$ GeV. Then the
Lorentz factor of muon is $\gamma=12.3$. The travel path of muon
in the laboratory frame is $L_0=810$ km.
For the distance between the accelerator and detector $L=250$ km
the fraction of muons undecayed at the Far Detector is
$k={\mathrm{exp}}(-L/L_0)={\mathrm{exp}}(-250/810)=0.73$.
This factor yields correction to the number of expected events
$158.1\times 0.73=115$ that is close to 112 observed events.

The MINOS experiment~\cite{Mi} detected muon neutrinos from the
Fermilab NuMI beam in two detectors located along the beam axis at
distances of 1 km (Near Detector) and 735 km (Far Detector).
It is measured the rate of $\nu_\mu$
born in the decay of $\pi^+$.
The neutrino energy distribution is centered on $E\sim 3.1$ GeV.
215 events are observed at the Far Detector compared to an
expectation of $336\pm 14.4$ events.

The deficit of muon neutrino flux in the MINOS experiment may be
accounted for by the entanglement of muon and neutrino
states likewise in the K2K experiment.
Estimate the effect for the peak neutrino energy $E\sim 3.1$ GeV
hence for the muon energy $E=3.1$ GeV. Then the
Lorentz factor of muon is $\gamma=29.2$. The travel path of muon
in the laboratory frame is $L_0=1\, 930$ km.
For the distance between the accelerator and detector $L=735$ km
the fraction of muons undecayed at the Far Detector is
$k={\mathrm{exp}}(-735/1\, 930)=0.68$.
This factor yields correction to the number of expected events
$336\times 0.68=228$ that is close to 215 observed events.

We have considered the entanglement of states of muon and muon
antineutrino born in the decay of charged pion. The decay of muon
with the left helicity triggers the left helicity for muon
antineutrino to preserve the null total angular momentum of muon
and muon antineutrino. This is forbidden for antineutrino hence one
cannot detect the muon antineutrino after the decay of muon.
This effect may explain the
deficit of muon neutrino flux in the Super-Kamiokande, K2K, MINOS
experiments. Now the deficit of muon neutrino flux in the
Super-Kamiokande~\cite{SK}, K2K~\cite{K2K}, MINOS~\cite{Mi}
experiments are explained with neutrino oscillations.
Non-detection of neutrinos due to the entanglement of muon and
neutrino
states is alternative to the effect of neutrino oscillations in
explanation of the deficit of muon neutrino flux in the
Super-Kamiokande, K2K, MINOS experiments.

\end{document}